\documentstyle[preprint,aps,epsfig]{revtex}
\newcommand{\lsim}{\mathrel{\mathop{\kern 0pt \rlap
  {\raise.2ex\hbox{$<$}}}
  \lower.9ex\hbox{\kern-.190em $\sim$}}}
\newcommand{\gsim}{\mathrel{\mathop{\kern 0pt \rlap
  {\raise.2ex\hbox{$>$}}}
  \lower.9ex\hbox{\kern-.190em $\sim$}}}
\newcommand{\es}     {\epsilon}
\newcommand{\R}     {{\mathcal R}}
\newcommand{\sh}     {\hat{s}}
\newcommand{\th}     {\hat{t}}
\newcommand{\uh}     {\hat{u}}
  \newcommand{\no}     {\nonumber}
  
  \newcommand{\Gm}     {\Gamma}
  \newcommand{\Mpl}     {M_{\rm Pl}}
\newcommand{\lm}     {\lambda}
\newcommand{\sg}     {\sigma}
\newcommand{\Lm}     {\Lambda}
\newcommand{\be}     {\begin{equation}}
\newcommand{\ee}     {\end{equation}}
\newcommand{\bea}     {\begin{eqnarray}}
\newcommand{\eea}     {\end{eqnarray}}
\begin{document}
\draft
\preprint{
\vbox{\hbox{hep-ph/0103308}
      \hbox{SNUTP 01-015}
      \hbox{KIAS P01018}}
}
\title{ $Z$ boson pair production at LHC \\
in a stabilized Randall-Sundrum scenario
}

\author{
Seong Chan Park$^{\:a}$,~~ H.~S.~ Song$^{\:a}$ and~~ Jeonghyeon~
Song$^{\:b}$ }
\vspace{1.5cm}
\address{
$^a$Department of Physics, Seoul National University,
Seoul 151-742, Korea\\
$^b$School of Physics,
Korea Institute for Advanced Study, Seoul 130-012, Korea
}

\maketitle
\thispagestyle{empty}

\setcounter{page}{1}

\begin{abstract}
\noindent
We study the $Z$ boson pair production at LHC in
the Randall-Sundrum scenario
with the Goldberger-Wise stabilization mechanism.
It is shown that comprehensive account of the Kaluza-Klein
graviton and radion effects is crucial to probe the model:
The KK graviton effects enhance
the cross section of $g g \to Z Z$ on the whole
so that the resonance peak of the radion
becomes easy to detect,
whereas the RS effects on the $q\bar{q} \to Z Z$ process
are rather insignificant.
The $p_T$ and invariant-mass distributions
are presented to study the dependence of
the RS model parameters.
The production of longitudinally polarized
$Z$ bosons,
to which the SM contributions are suppressed,
is mainly due to KK gravitons and the radion,
providing one of the
most robust methods to signal the RS effects.
The $1 \sigma$ sensitivity bounds on $(\Lambda_\pi, m_\phi)$
with $k/M_{\rm Pl} =0.1$ are also obtained such that
the effective weak scale $\Lambda_\pi$ of
order $5$ TeV can be experimentally probed.
\end{abstract}
\newpage

\section{Introduction}

Recent advances in string theories
have inspired particle physicists
to approach the gauge hierarchy problem
of the standard model (SM) in a very novel way,
i.e., by introducing extra dimensions.
Arkani-Hamed, Dimopoulos and
Dvali (ADD) proposed that
there exist $n$ large extra dimensions
with factorizable geometry,
whereas the
SM fields are confined to our four-dimensional
world{\cite{Antoniadis:1998ig}}.
The observed largeness of Planck scale $M_{\rm Pl}$
is attributed to the large volume of the extra
dimensions $V_n$,
as can be seen from the relation $M_{\rm Pl}^2=M_S^{n+2} V_n$
with $M_S$ being the fundamental scale.
Since the $M_S$ can be maintained around TeV scale,
the hierarchy problem is answered.
Criticism arose such that the ordinary gauge hierarchy is
replaced by a new hierarchy between the $M_S$ and
the compactification scale $\mu_c=V_n^{-1/n}$.
Based on two branes and a single extra dimension
with non-factorizable geometry,
Randall and Sundrum (RS) have proposed another
higher dimensional scenario where,
without $large$ volume of the extra dimension,
the hierarchy problem is solved by a geometrical
exponential factor{\cite{Randall:1999ee}}.
Here the stabilization of the compactification
radius is very crucial:
Otherwise the study of the cosmological evolution
in the RS scenario has shown that
a nontrivial relationship
between the matter densities on the two
branes is required, which induces non-conventional
cosmologies \cite{Csaki:2000mp}.
Goldberger and Wise (GW)
have proposed a stabilization mechanism
where a bulk scalar field
with interactions localized on the two branes
can generate
for the modulus field a potential
which allows a minimum appropriate to the
hierarchy problem \cite{Goldberger:1999uk}.

Of great interest and significance
is that the ADD and RS models could have
chances to be detected at future collider
experiments.
Even more interestingly,
they provide possible accounts for the recently
reported deviation of the muon anomalous
magnetic moment from the SM prediction \cite{Kim:2001rc}.
In the ADD case,
even though the coupling
of each Kaluza-Klein (KK) graviton state
with the SM fields is suppressed by Planck scale,
summation over almost continuous KK spectrum
compensates the suppression and leaves
the effective coupling of
$\sim 1/M_S$ \cite{Wells}.
In the RS scenario,
the zero mode of the KK graviton states
couples with the usual Planck strength, whereas
the masses and couplings of all the
excited KK states are characterized by some
electroweak scale $\Lm_\pi$
\cite{Davoudiasl:2000jd,Goldberger:1999wh}.
This discrete spectrum is to
yield the clean signal of
graviton resonance production.
Another key ingredient of the RS model is
from the stabilization mechanism, the radion.
Since
the radion can be much lighter than
$\Lm_\pi$ \cite{Giudice:2000av,Ko},
it is likely that the first
signal of the RS effects come from the radion.

Various phenomenological aspects of the radion
have been extensively studied in the literature:
The decay modes of the radion
are different from those of the Higgs boson
(e.g., the radion with mass smaller $2 m_Z$
dominantly decays into two gluons) \cite{Giudice:2000av,Ko};
without a curvature-scalar Higgs mixing,
the radion effects
on the oblique parameters
of the electroweak precision observations are small
\cite{Csaki:2001zn};
the radion effects on the phenomenology at low energy
colliders \cite{Mahanta:2000zp}
and at high energy colliders
\cite{Mahanta:2000ci,Choudhury:2000fj,Cheung:2001rw}
have been also discussed.

In the RS scenario, however,
there is another key ingredient,
the KK spectrum of gravitons.
Especially at high energy collider,
comprehensive effects of the radion and KK gravitons
can be substantial.
Moreover, in spite of its lighter mass
than the KK gravitons,
the coupling strength of the radion with the SM particles
is weaker than that of the KK gravitons
due to the following reasons:
The characteristic scale of the radion
coupling, inversely proportional to the vacuum expectation
value (VEV) of the radion
($\Lm_\phi$), is smaller than the coupling of the KK gravitons,
since $\Lm_\phi=\sqrt{6}\,\Lm_\pi$ \cite{Giudice:2000av};
the degrees of freedom of the spin-0 radion
are less than those of the spin-2 massive KK gravitons.
And the $\Lm_\phi$,
usually treated as a free parameter,
is also constrained through the relation with the $\Lm_\pi$
which receives various constraints
from the LEP II and Tevatron experiments
\cite{Davoudiasl:2000jd}.
It is to be shown that
the comprehensive accommodation at high energy colliders
leads to different phenomenologies
from those only with the radion effects.

In order to sensitively probe the radion effects,
we consider the process $g g \to Z Z$.
Note that the radion interacts with the SM fields through
the trace of the energy-momentum tensor:
The coupling becomes stronger
as the interacting SM particles are more massive;
QCD trace anomaly enhances the coupling
of the radion to a pair of gluons.
This process has been extensively studied with
special attention to the Higgs search at the LHC
\cite{Glover:1989rg}.
The double leptonic decay of the $Z$-boson,
$g g \to ZZ \to l^+ l^- l'^+ l'^-$,
generates a clean signal.
Unfortunately
the main background of the continuum production of
the $Z$ pair via $q\bar{q}$ annihilation
is known to be dominant
except for a limited range of the
Higgs resonance.
For example, full one-loop calculations for the
$g g \to Z Z$
in the minimal supersymmetric standard model (MSSM)
with the squark loop contributions
have shown that the irreducible background of
$q\bar{q}\to Z Z$ surpasses
even resonant peaks
of the supersymmetric Higgs bosons\cite{Berger:1999vx}.

As shall be shown,
comprehensive accommodation of both the KK graviton
and radion effects is very crucial to
explore the RS model.
The RS effects on the
gluon fusion process are much larger than
those on the $q\bar{q}$ process,
which shall be apparently shown in the
$p_T$ and invariant-mass distributions.
As a result,
the radion effects on the $g g \to Z Z$
have much more chance to be detected,
while the radion effects on the $q\bar{q} \to Z Z$
are negligible due to the smallness of $m_q$.
Moreover the measurement of the $Z$ polarization
shall provide another efficient method for probing the
effects of the RS model.

This paper is organized as follows.
In Sec.~II, we briefly review the RS model
with the GW mechanism, and
summarize the effective Lagrangian
between the KK graviton, radion
and SM particles.
Model parameters are carefully notified.
In Sec.~III,
the parton level helicity amplitudes
for $gg \rightarrow ZZ$ and $q \bar{q} \rightarrow ZZ$
to leading order
are given.
In Sec.~IV, we present numerical results of
the $p_T$ and invariant-mass distributions.
After presenting the RS model dependence
on the distributions,
we shall show the RS effects on
various configurations of $Z$-boson polarization.
The sensitivity bounds for parameter space of
$\Lambda_\pi$ and the radion mass are to be given.
Finally, Sec.~V deals with summary and conclusions.

\section{Stabilized Randall-Sundrum scenario}
For the hierarchy problem,
Randall and Sundrum have proposed a five-dimensional
non-factorizable geometry with the extra dimension
compactified on a $S_1/Z_2$ orbifold of radius $r_c$.
Reconciled with four-dimensional Poincar${\rm \acute{e}}$ invariance,
the RS configuration has the following solution to
Einstein's equations:
\begin{equation}
\label{metric}
d s^2= e^{-2 k r_c |\varphi| } \eta_{\mu\nu} d x^\mu d x^\nu
+ r_c^2 d \varphi^2,
\end{equation}
where $0 \leq |\varphi| \leq \pi$
and $k$ is $ AdS_5$
curvature.
Two orbifold fixed points accommodate two three-branes,
the hidden brane at $\varphi=0$ and
our visible brane at $|\varphi|=\pi$.
The arrangement of our brane at $|\varphi|=\pi$
renders a fundamental scale $m_0$ to appear as
the four-dimensional physical mass $m=e^{-k r_c \pi} m_0$.
The hierarchy problem can be answered
if $k r_c \approx 12$.
From the four-dimensional effective action,
the relation between the four-dimensional Planck scale
$M_{\rm Pl}$
and the fundamental string scale $M_S$
is obtained by
\be
\label{Mpl}
M_{\rm Pl}^2=\frac{M_S^3}{k}(1-e^{-2kr_c \pi}).
\ee

The compactification of the fifth dimension
leads to the following four-dimensional effective Lagrangian
\cite{Davoudiasl:2000jd},
\begin{equation}
\label{Lgraviton}
{\mathcal L} = -\frac{1}{M_{\rm Pl}} T^{\mu\nu}
h^{(0)}_{\mu\nu}
-\frac{1}{\Lambda_\pi}  T^{\mu\nu} \sum_{n=1}^\infty
h^{(n)}_{\mu\nu}
\,,
\end{equation}
where $ \Lambda_\pi \equiv  e^{-k r_c \pi} M_{\rm Pl}$
is at the electroweak scale.
The coupling of the zero mode of the KK gravitons
is suppressed by usual Planck scale,
while those of the massive KK gravitons
by the electroweak scale
$\Lambda_\pi$.
The masses of the KK gravitons are also at the electroweak scale,
given by \cite{Davoudiasl:2000jd,Goldberger:1999wh},
\begin{equation}
\label{Gmass}
m_n=k x_n e^{-k r_c \pi}=\frac{k}{M_{\rm Pl}} {\Lambda_\pi} x_n
\,,
\end{equation}
where the $x_n$'s are the $n$-th roots
of the first order Bessel function.
The condition $k<M_{\rm Pl}$ is to be imposed
to maintain the reliability of the RS solution
in Eq.~(\ref{metric}) \cite{Davoudiasl:2000tf}.
We take the value in the conservative
range of $0.1<k/M_{\rm Pl}<0.7$.
Then, the first excited
KK graviton has mass slightly larger than $1$ TeV for
$\Lambda_\pi \sim 3$ TeV and so
there might be a chance to see the effects of KK gravitons
at future high energy colliders.

In the original RS scenario, the compactification radius
$r_c$ is assumed to be constant.
According to the studies of cosmological evolution,
however,
two branes wants to blow apart, i.e., $r_c \to \infty$,
unless we impose a fine-tuning between the densities
on two branes, which will lead to
non-conventional cosmologies \cite{Csaki:2000mp}.
A stabilization mechanism is required.
Goldberger and Wise have introduced a bulk scale field
with the bulk mass somewhat smaller than the $k$.
The assumption of localizing
the bulk scalar interactions
on two branes determines the four-dimensional
effective potential
for the radion, which can allow the minimum for
$k r_c \approx 12 $ without an extreme  fine-tuning.
Furthermore, the radion mass is
roughly an order of magnitude below $\Lm_\pi$
\cite{Csaki:2001zn}.

It was shown that the radion couples to ordinary matter
through the trace of the symmetric and conserved
energy-momentum tensor with TeV scale suppressed
coupling;
\be
\label{Lradion}
{\cal L}=\frac{1}{\Lambda_\phi}\,\phi \,T_\mu^\mu,
\ee
where $\Lm_\phi$, the VEV of
the radion field,
is related such that
$\Lambda_\phi=\sqrt{6}\,\Lambda_\pi$ \cite{Giudice:2000av}.
The coupling of the radion with a fermion or
massive gauge boson pair
is the same as that of the Higgs,
except for a factor of $(v/\Lm_\phi)$,
with $v$ being the VEV of the SM Higgs boson.
The massless gluons and photons also contribute
to the $T^\mu_\mu$,
due to the trace anomaly which
appears
as the scale invariance of massless fields
is broken by the running of gauge couplings\cite{Collins:1977yq}.
Thus the interaction Lagrangian between two gluons and
the radion or Higgs boson is
\begin{equation}
{\mathcal L}_{h(\phi)-g-g}
=
\left[
\left( \frac{v}{\Lm_\phi} \right)
\left\{
b_3 + I_{1/2}(z_t^\phi)
\right\} \phi
+
I_{1/2}(z_t^h) h
\right]
\frac{\alpha_s}{8 \pi v} \,{\rm Tr}
(G^C_{\mu\nu}G^{C \mu\nu} )
\,,
\end{equation}
where $z_t^{\phi (h)} = 4 m_t^2/m_{\phi (h)}^2$,
$m_t$ is the top quark mass,
and
the QCD beta function coefficient is
$b_3 =11-2 n_f /3$ with the number of dynamical quarks $n_f$.
The loop function $ I_{1/2}(z) $ is defined by
\be
I_{1/2}(z) =z [ 1+ (1-z) f(z)]
\,,
\ee
where the $f(z)$ is
\begin{eqnarray}
\label{ffnt}
f(z) &=& \left\{  \begin{array}{cl}
           {\rm \arcsin}^2(1/\sqrt{z}) \,,
           &\qquad z \geq 1\,, \\
           -\frac{1}{4}\left[\ln
           \left(\frac{1+\sqrt{1-z}}{
1-\sqrt{1-z}}\right)
                             -i\pi\right]^2\,,
                             & \qquad z < 1\,.
               \end{array}
\right.
\end{eqnarray}
It is to be noted that
the phenomenology of radions and KK gravitons in the RS model
can be determined by three parameters,
$m_\phi$, $\Lm_\pi$, and $k/M_{\rm Pl}$.

\section{Z boson pair production at LHC}

\subsection{The $g g \to Z Z$ helicity amplitudes}

For the process
$g(\lambda_1) g(\lambda_2)
\rightarrow Z(\lambda_3) Z(\lambda_4)$
there are, in general,  36 helicity amplitudes according to
two and three polarization states of initial gluons and final
$Z$-bosons, respectively.
Various symmetry arguments reduce these 36 amplitudes
into eight independent ones \cite{Glover:1989rg}.
Parity invariance implies
\be
{\cal M}_{\lambda_1\lambda_2 \lambda_3 \lambda_4}
={\cal M}^*_{-\lambda_1 -\lambda_2 -\lambda_3 -\lambda_4}.
\ee
Bose statistics and the standard form of
the $Z$-boson polarization vectors demand
\bea
{\cal M}_{++--}(\beta) &=&{\cal M}_{++++}(-\beta),\\
{\cal M}_{+++-}(\beta) &=&{\cal M}_{++-+}(\beta),\\
{\cal M}_{+---}(\beta) &=&{\cal M}_{+-++}(\beta),\\
{\cal M}_{+--+}(\beta) &=&{\cal M}_{+-+-}(-\beta),\\
{\cal M}_{+++0}(\beta) ={\cal M}_{++0+}(\beta)
&=&{\cal M}_{++-0}(-\beta)={\cal M}_{++0-}(-\beta),\\
{\cal M}_{+-+0}(\beta) =-{\cal M}_{+-0+}(-\beta)
&=&{\cal M}_{+--0}(-\beta) =-{\cal M}_{+-0-}(\beta).
\eea

For the explicit calculations, we consider the following
four momenta for the initial gluons and final $Z$ bosons
in the gluon-gluon center-of-momentum (c.m.)
frame:
\bea
p_1= \frac{\sqrt{\sh}}{2}(1,0,0,1),
&\quad&
 p_2= \frac{\sqrt{\sh}}{2}(1,0,0,-1), \\ \no
p_3 = \frac{\sqrt{\sh}}{2}
(1, \beta \sin \theta , 0 , \beta \cos \theta),
&\quad&
p_4 =\frac{\sqrt{\sh}}{2}
(1,- \beta \sin \theta , 0 ,- \beta \cos \theta),
\eea
where $\beta=\sqrt{1-{4 m_Z^2}/{\sh}}$ and $\th=(p_1-p_3)^2$.
The polarization vectors for the spin-1 particles
with momentum $p^\mu =(p^0, \vec{p}\,)$ are
\bea
\epsilon^\mu (p,\lambda)&=&
\frac{e^{i \lambda \phi_{\hat{p}}}}{\sqrt{2}}\,
(0, -\lambda \cos \theta \cos \phi_{\hat{p}} +
i \sin \phi_{\hat{p}}, -i \cos \phi_{\hat{p}}
 -\lambda \cos \theta
 \sin \phi_{\hat{p}} , \lambda \sin \theta ),\\
\epsilon^\mu (p, 0) &=&~~~~~~
({\left| \vec p \,\right|}/{m},\,
{p^0}\hat{p}/m),
\eea
where the angles $\theta$ and
$\phi_{\hat{p}}$ specify the direction of $\vec{p}$ .

\begin{center}
\begin{figure}[htb]
\vspace{\bigskipamount}
\hbox
to\textwidth{\hss\epsfig{file=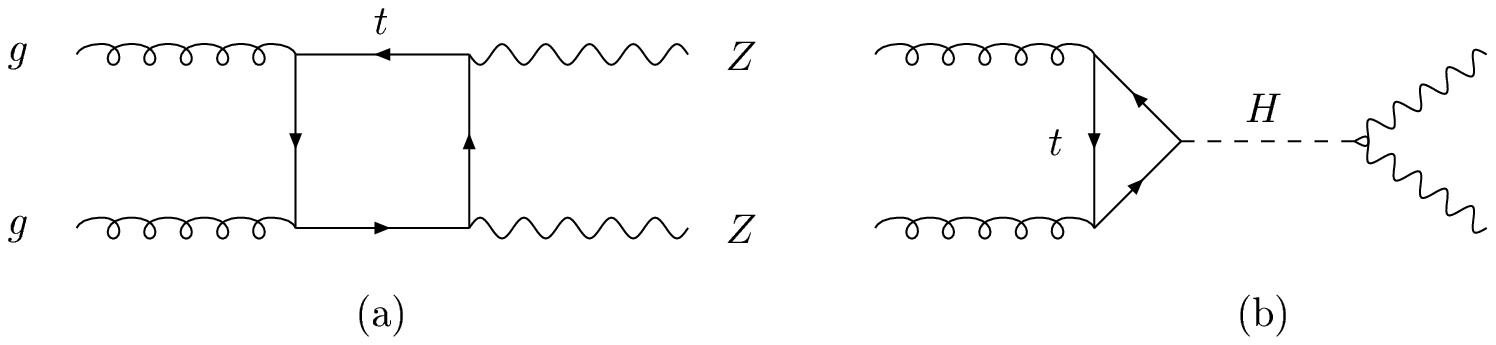,scale=0.7}\hss}
\smallskip
\smallskip
\smallskip
\caption{ Feynman diagrams for the process
$g g\to Z Z$ in the SM. } \label{fig1}
\end{figure}
\end{center}

In the SM, there are two types of Feynman diagrams
for the $g g \to Z Z$ process, through the box and triangle
quark loops shown in Fig.~\ref{fig1}.
Despite loop suppression by a factor of $\alpha_s^2$,
high luminosity of the gluon in a proton at the LHC
yields substantial cross section for the process.
The SM helicity amplitudes have been studied in detail.
We refer the reader to Ref.~\cite{Glover:1989rg}.

\begin{center}
\begin{figure}[htb]
\hbox
to\textwidth{\hss\epsfig{file=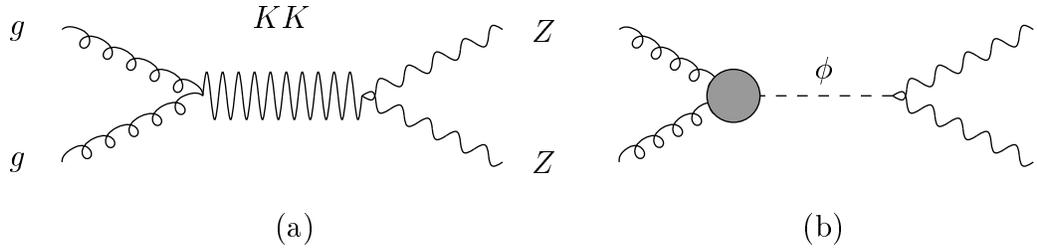,scale=1}\hss}
\smallskip
\smallskip
\smallskip
\caption{Feynman diagrams
for the process $g g \to Z Z$ mediated (a) KK gravitons
and (b) the radion in the RS model. } \label{fig2}
\end{figure}
\end{center}

In the RS model, KK gravitons mediate the $s$-channel
Feynman diagram at tree level (see Fig.~\ref{fig2}).
The helicity amplitudes are cast into
\be
{\cal M}^{G} _{\lambda_1 \lambda_2 \lambda_3 \lambda_4}
( g g \to Z Z)
=- \frac{1}{8 \Lambda_\pi^2} \sum_n
\frac{\delta^{ab}}{\sh -m_n^2}
{\cal A}_{\lambda_1 \lambda_2 \lambda_3  \lambda_4}
\,,
\ee
where $\delta^{a b}$ denotes the color factor.
There are six non-vanishing independent helicity amplitudes:
\bea
{\cal A}_{++++}&=& -\frac{1}{2}( \beta^4-1)(\th-\uh)^2
+\frac{1}{2}(\beta^2+1)\sh^2
                 + (\th+\uh)\sh , \\
{\cal A}_{+-+-}&=&\frac{1}{2}\{\beta (\th-\uh) +\sh \}^2 ,\\
{\cal A}_{+-++}&=&\frac{1}{2}(\beta^2-1)\{\beta^2 (\th-\uh)^2 -\sh^2 \}  , \\
{\cal A}_{+-+0}&=&-\frac{\sqrt{\sh}}{2\sqrt{2}m_Z}
(1- {1}/{\beta^2})(\th -\uh +\beta \sh)
               \sqrt{(\beta \sh)^2 -(\th-\uh)^2  }, \\ \no
{\cal A}_{++00}&=&\frac{(1+\beta^2)
\{ (1+\beta^2)\sh +2 (\th+\uh)\}\sh^2}
                     {8 m_Z^2} ,\\
{\cal A}_{+-00}&=&\frac{(1-{1}/{\beta^2})(\beta^2-2)
\{(\th-\uh)^2-\beta^2\sh^2\}
                \sh}{8 m_Z^2}.
\eea

The second ingredient of the RS model, the radion,
couples to two gluons
through its Yukawa coupling to a quark inside
a triangle loop diagram,
as well as through QCD trace anomaly.
Since the radion interaction with quarks
is proportional to the mass, only top quark loop
is to be considered.
Two of the eight independent helicity amplitudes
receive non-vanishing contributions
from the radion:
\bea
{\cal M}^\phi_{++++}&=&
-\frac{\alpha_s \delta^{ab}}{\pi}[b_{QCD}+ I(z_t)]
\left(\frac{m_Z}{\Lambda_\phi}\right)^2
\frac{1}{\sh-m_\phi^2+i\Gamma_\phi m_\phi }\frac{\sh}{2},\\ \no
{\cal M}^\phi_{++00}&=&-\frac{\alpha_s\delta^{ab}}{\pi}
[b_{QCD}+ I(z_t)]
\left(\frac{m_Z}{\Lambda_\phi}\right)^2
\frac{1}{\sh-m_\phi^2+i\Gamma_\phi m_\phi }
\frac{\sh^2(\beta^2+1)}{8 m_Z^2}.
\eea

\subsection{The $q\bar{q} \to Z Z$ helicity amplitudes}

In the SM the process $ q\bar{q} \to Z Z$
surpass the gluon fusion process
due to the presence of tree level Feynman diagrams.
Accurate structure functions at low $x$
and higher order QCD corrections are necessary
for the precise ratio of the
$q\bar{q}$ annihilation and gluon fusion processes.
Even though the $K$-factor of ${\mathcal O}(30\%)$
of the
$ q\bar{q} \to Z Z$ is known
 \cite{Barger:1988fm},
the absence of the corresponding calculation for
the gluon fusion process
leads us not to include any higher order corrections.
We also assume that the uncertainties
with leading contributions from soft gluon emission
in both processes cancel to some extent
in the ratio of cross sections,
hardly affecting our main interest, the distribution shapes.

In the SM, there are $t$- and $u$-channel Feynman diagrams.
Since the radion coupling to fermions
is proportional to the fermion mass, we can
safely neglect the radion effects here.
The $s$-channel diagram mediated by
KK gravitons,
which is similar to the diagram $(a)$ in Fig.~2,
still influence the production.
For the
process $q(\lambda_1) \bar{q} (\lambda_2 ) \rightarrow
         Z(\lambda_3) Z(\lambda_4) $,
the helicity amplitudes due to KK gravitons are
written by
\be
 {\cal M}^{G}_{\lambda_1 \lambda_2 \lambda_3 \lambda_4}
 (q \bar{q} \to Z Z)
 = - \frac{1}{4 \Lambda_\pi^2}
 \sum_n \frac{\delta^{\alpha \beta}}{\hat{s}-m_n^2}
 {\cal B}_{\lambda_1 \lambda_2 \lambda_3 \lambda_4},
\ee
where non-zero
and independent ${\cal B}$'s are, in the parton c.m. frame,
\bea
{\cal B}_{+-++}&=&
\frac{\sh(\th-\uh)(\beta^2-1)\sin\theta}{\beta},\\\no
{\cal B}_{+-+-}&=&
\frac{\sh (\th-\uh+\beta \sh) \sin\theta }{\beta}, \\ \no
{\cal B}_{+-+0}&=&
-\frac{(\beta^2 -1)\sh^{3/2}
(1+\cos\theta)\{\beta \sh - 2(\th-\uh)\}}{2^{3/2} m_Z \beta },\\ \no
{\cal B}_{+--+}&=&
\sh^2 \sin\theta (\cos\theta-1), \\ \no
{\cal B}_{+--0}&=&
-\frac {(\beta^2 -1)\sh^{3/2}(1-\cos\theta)
(\beta \sh + 2(\th-\uh))}{2^{3/2} m_Z \beta },\\ \no
{\cal B}_{+-00}&=&
\frac{\sh^2 (\beta^2-1)(\beta^2-2)(\th-\uh)\sin\theta}{4m_Z^2 \beta},
\eea
where $\cos\theta=(\th-\uh)/\beta\sh$.

\section{Continuum $Z$ boson pair production at the LHC}
The physical production rate of $Z$ boson pair
at $p p$ colliders
is obtained by convoluting
over the parton structure functions \cite{Eichten:1984eu}:
\be
 \sigma (pp \rightarrow ij \rightarrow ZZ)
= \int dx_1 dx_2
\sum_{i,j}
f_i (x_1, Q^2) f_j (x_2, Q^2) \hat{\sigma}_{ij}(x_1 x_2 s )
\,,
\ee
where $i$ and $j$ denote the parton
such as the gluon or quark,  and the $x_1$ and
$x_2$ denote the momentum fraction
of the parton from the parent
proton beam.
For the numerical
analysis, we use the leading order MRST parton distribution
functions (PDF)\cite{MRST}. The QCD
factorization and renormalization scales $Q$ are set to be the
$m_Z$.
The $Q^2$-dependence is expected to be small on
the distribution shapes.
The c.m. energy at the LHC
is $\sqrt{s}=14$ TeV.
And we have employed the
kinematic cuts of $p_{_T} \geq 25$ GeV and
$|\eta| \leq 2.5$ throughout
the paper.
As discussed before,
the RS effects are determined by
three parameters,
 $(\Lm_\pi, {k}/M_{\rm Pl} , m_\phi)$.
From the above arguments,
we consider $\Lambda_\pi = 2,~3,
~5$  TeV,
$k/M_{\rm Pl} = 0.1 ,~ 0.3,
~0.7$,
and $m_\phi = 300,~ 500, ~700$ GeV.

\begin{center}
\begin{figure}[htb]
\hbox
to\textwidth{\hss\epsfig{file=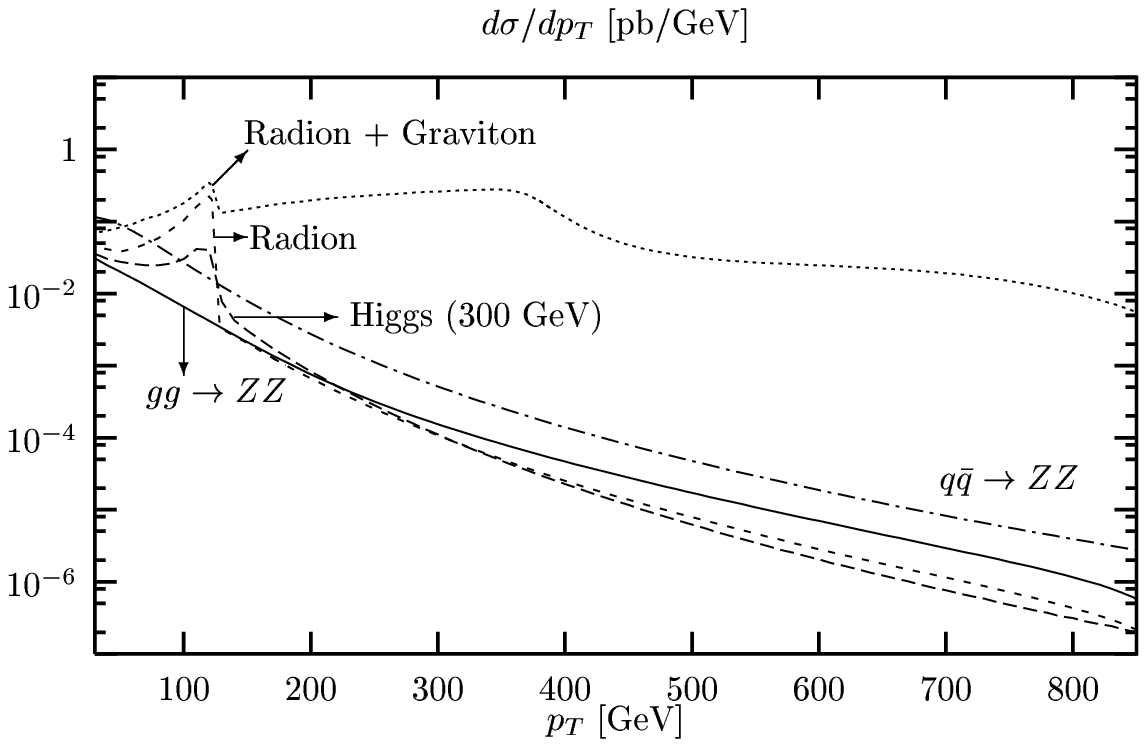,scale=1.1}\hss}
\bigskip
\hbox
to\textwidth{\hss\epsfig{file=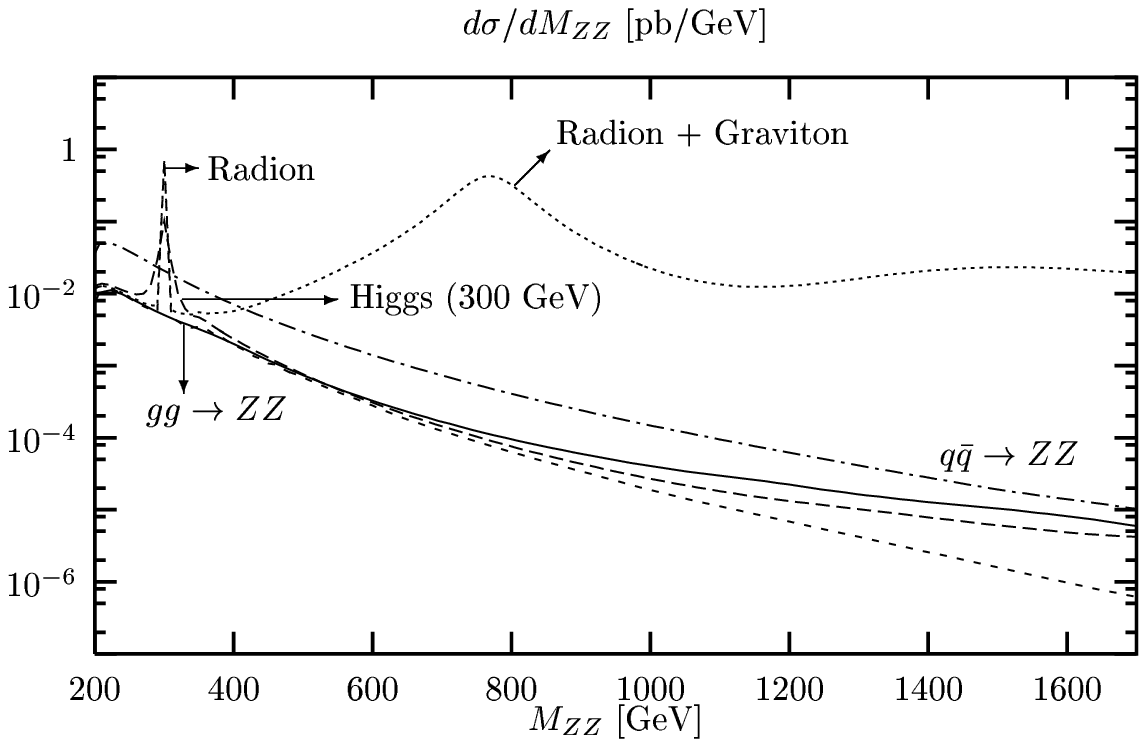,scale=1.1}\hss}
\smallskip
\smallskip
\smallskip
\caption{The $p_T$ and invariant-mass distributions
of the $g g\to ZZ$ process
when both the SM Higgs boson and radion masses are 300 GeV
with $\Lm_\pi=2$ TeV and $k/M_{\rm Pl}=0.1$.
The SM results for $q\bar{q}\to Z Z$ are plotted
for comparison.} \label{hmrm}
\end{figure}
\end{center}

The SM Higgs boson mass has not been experimentally confirmed yet.
Recently, the ALEPH group has reported the observation
of an excess of $3\,\sigma$ in the search of the SM Higgs boson,
which corresponds to the Higgs mass about 114 GeV\cite{ALEPH}.
As the operation of LEP II has been completed,
the decision whether the observations are only the results of statistical
fluctuations or the first signal of the Higgs boson production
is suspended until the Tevatron II
and/or LHC running \cite{Ellis}.
In the following,
the Higgs boson mass is set to be
114 GeV except for the comparison
of the contributions from the Higgs boson and radion
with the same mass in Fig.~\ref{hmrm}.

Before presenting numerical results,
some discussions on the unitarity violation
of the RS model are in order here.
As can be seen in the effective Lagrangian of
Eqs.~(\ref{Lgraviton}) and (\ref{Lradion}),
the RS model generically undergoes the unitarity
violation at high energies $\sqrt{\hat{s}} \gg \Lm_\pi$.
In Ref.~\cite{Song},
the elastic process
$\gamma \gamma \to \gamma\gamma$ has been examined
to obtain the bound from partial wave unitarity
on the ratio $\sqrt{\sh}/ \Lm_\pi$ in the RS model.
This process can yield very sensitive bounds since
the RS effects mediated by KK gravitons are dominant
due to the absence of the SM contributions.
The $J$-partial wave amplitude is defined \cite{Eboli:2000aq} by
\begin{equation}
a^J_{\mu\mu^\prime} = \frac{1}{64 \pi}
\int^1_{-1} d\cos\theta\; d^J_{\mu\mu^\prime}(\cos\theta)
\,\left[
-i{\cal M}_{\lambda_1\lambda_2\lambda_3\lambda_4}\right] \; ,
\label{aj}
\end{equation}
where
the ${\cal M}_{\lm_1\lm_2\lm_3\lm_4}$
is the helicity amplitudes,
$\mu =\lambda_1-\lambda_2$,
$\mu^\prime = \lambda_3-\lambda_4$,
and the $d^J_{\mu\mu^\prime}$ is the Wigner functions\cite{pdg}.
Unitarity implies that
the largest
eigenvalue ($\chi$)
of $a^J_{\mu\mu^\prime}$
is to be $|\chi| \le 1$.
The reliability of perturbative calculations
is approximately guaranteed
by the conditions
$|\chi|= 1$ and $|{\rm Re}(\chi )| = 1/2$.
The helicity amplitudes,
of which the dominant contribution at high energies
comes from the KK gravitons,
are
\begin{eqnarray}
{\cal M}_{++++} &=& {\cal M}_{----}
= -i~\frac{s^2}{\Lm_\pi^2}~
\sum_{n}\left[
D_n(t)
+
D_n(u)
\right]\; , \\ \no
{\cal M}_{+-+-} &=& {\cal M}_{-+-+}
= -i~\frac{u^2}{\Lm_\pi^2}~
\sum_{n}\left[
D_n(s)
+
D_n(t)
\right]\;, \\ \no
{\cal M}_{+--+} &=& {\cal M}_{-++-}
= -i~\frac{t^2}{\Lm_\pi^2}~
\sum_{n}\left[
D_n(s)
+
D_n(u)
\right]\;,
\end{eqnarray}
where $D_n(s)=1/(s-m_n^2+i m_n \Gm_n)$.
The odd $J$-partial wave amplitudes vanish
due to Bose-Einstein statistics
in the elastic $\gamma \gamma$ scattering.
And we have $a^2_{22} = a^2_{-2-2}$ and
$a^2_{-22} = a^2_{2-2}$ from the parity
arguments.
The non-vanishing  eigenvalues $\chi_i$ are
$a^2_{00}$ and $2 a^2_{22}$.
Numerical estimation leads to
$\sqrt{s} \lsim 3.1 \,\Lm_\pi$ for $k/\Mpl =0.1$,
$\sqrt{s} \lsim 5.7 \,\Lm_\pi$ for $k/\Mpl =0.3$, and
$\sqrt{s} \lsim 9.8 \,\Lm_\pi$ for $k/\Mpl =0.7$.
In what follows,
a very conservative bound
is to be employed such as $\sqrt{\hat{s}} \leq 0.9 \,\Lm_\pi$.
In order to eliminate the concern for unitarity violation,
it is reasonable that we restrict ourselves to
the region where our perturbative calculations
are trustworthy.
This can be achieved by excluding data with
high invariant-mass.
In the following, therefore,
we constrain the invariant mass to be less than $1.8$ TeV,
to be consistent with our parameterizations of $\Lm_\pi$.

\begin{center}
\begin{figure}[htb]
\hbox
to\textwidth{\hss\epsfig{file=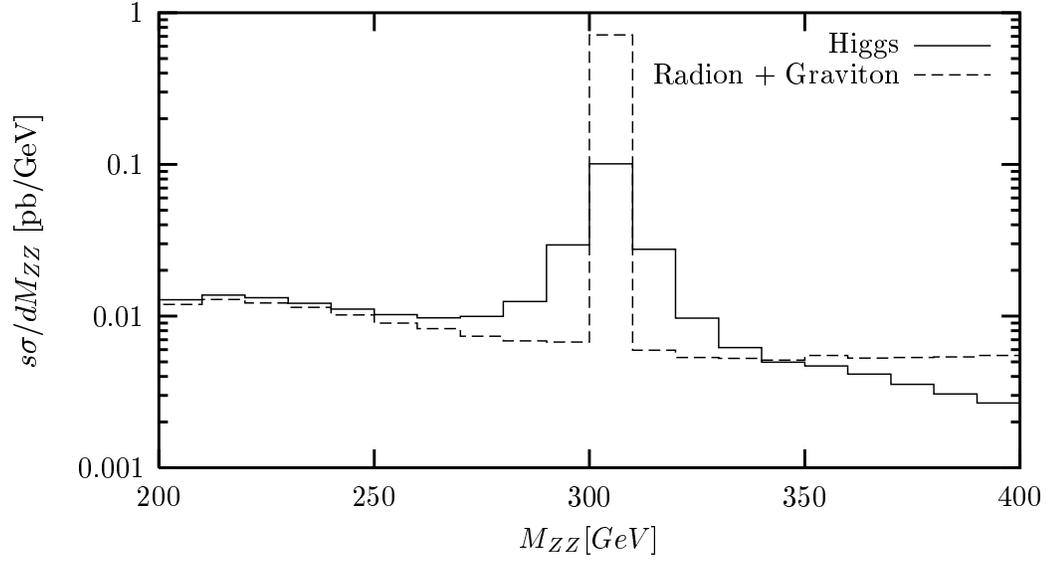,scale=1.1}\hss}
\smallskip
\caption{The resonance peak shapes
in the invariant-mass distributions
of the $g g\to ZZ$ process
when both the SM Higgs boson and radion masses are 300 GeV
with $\Lm_\pi=2$ TeV and $k/M_{\rm Pl}=0.1$.} \label{binM}
\end{figure}
\end{center}

First we present the $p_T$ and invariant-mass distributions
for unpolarized $Z$-bosons.
Figure \ref{hmrm} shows the radion and KK graviton effects
for the process $g g\to Z Z$
on the distributions.
For comparison, both the Higgs boson and radion masses
are set to be 300 GeV with
$\Lm_\pi=2$ TeV and $k/M_{\rm Pl}=0.1$.
The long dashed line denotes the SM results with
the Higgs mass of 300 GeV.
The short dashed line includes only the radion effects,
whereas the dotted line incorporates both
the KK graviton and radion effects.
The SM result of $q\bar{q} \to Z Z$
denoted by the dash-dotted line is also plotted
for comparison.
The $p_T$ distributions
apparently show that the KK graviton
effects enhance the cross section on the whole,
which increases the chance to probe the presence of the radion.
Any other models for new physics beyond the SM,
including supersymmetric models,
hardly generate such elevated resonance behavior.
Even in the large $\Lm_\pi$ cases
where the KK graviton effects are negligible,
the distinction between the Higgs boson and
radion is, in principle, possible:
The resonance peak of the radion
becomes narrower with increasing $\Lm_\pi$
because the radion total decay width
$(\Gamma_\phi)$
is inversely proportional to
$ \Lm_\pi^2$.
Figure \ref{binM} shows the resonance shapes of the
Higgs boson and radion with the same mass,
through the invariant mass spectrum of the $Z$ pair.
It can be seen that if both have the same mass,
the radion shows sharper resonance peak than
the Higgs boson.
This is generic
since the total decay width of the radion
is smaller than that of the Higgs boson
due to the radion's larger VEV.

\begin{center}
\begin{figure}[htb]
\hbox
to\textwidth{\hss\epsfig{file=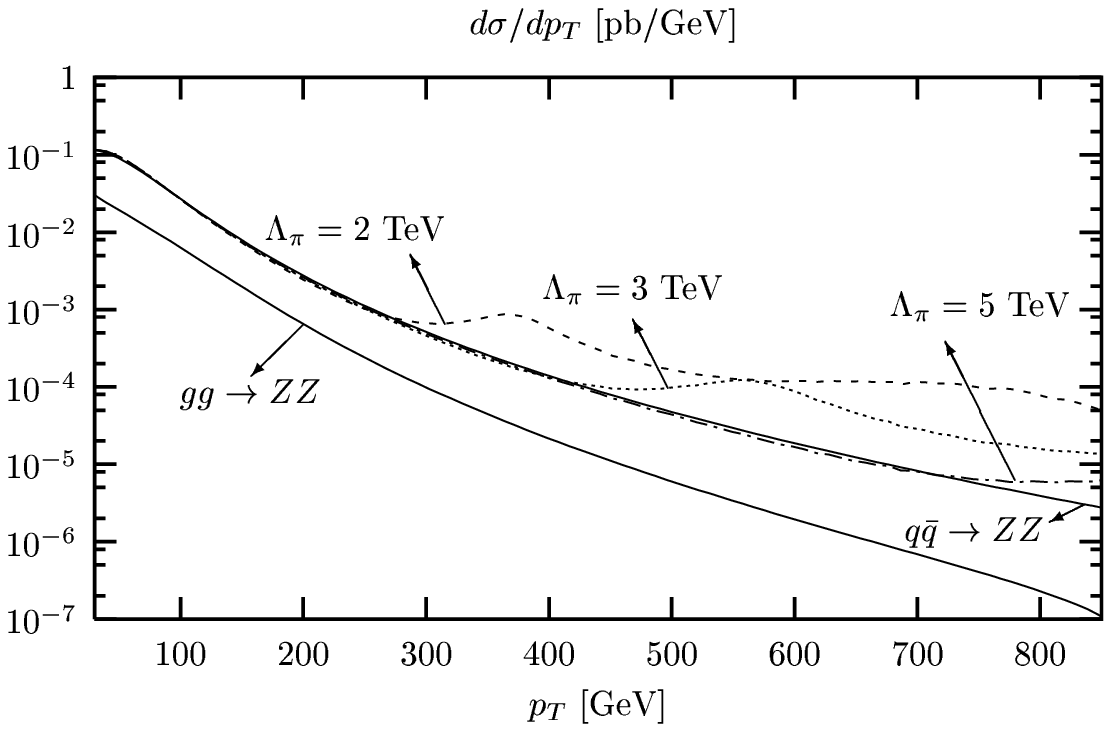,scale=1.1}\hss}
\bigskip
\hbox
to\textwidth{\hss\epsfig{file=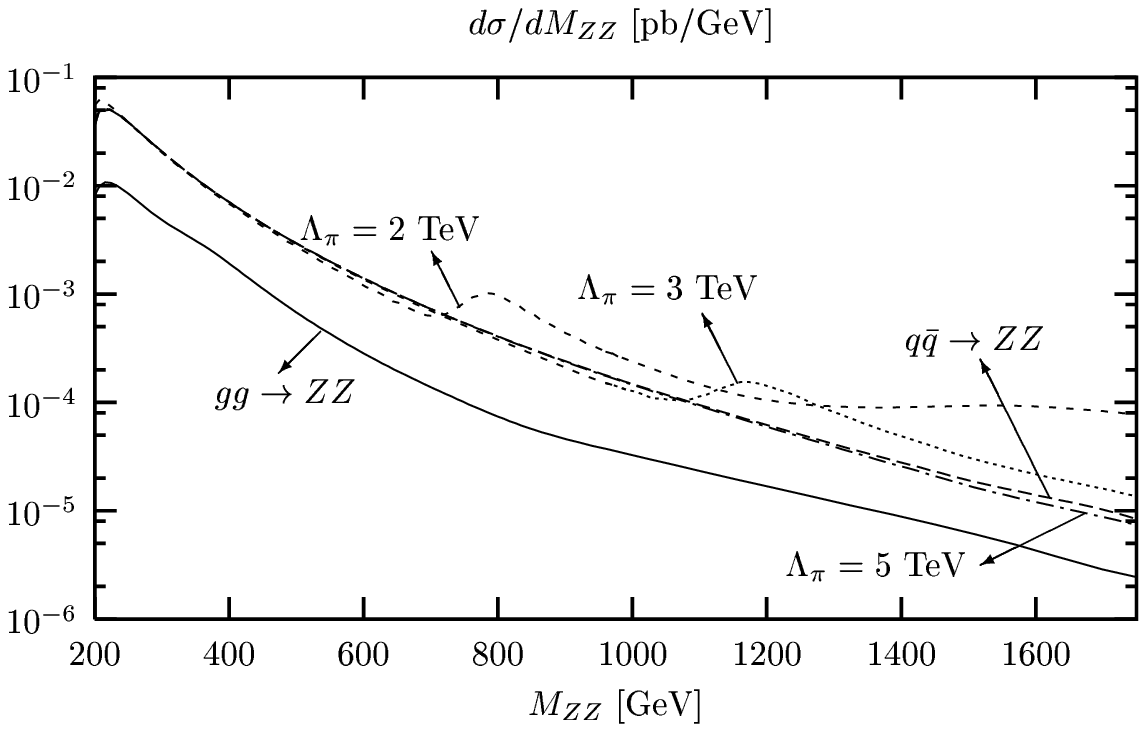,scale=1.1}\hss}
\smallskip
\smallskip
\smallskip
\caption{The $p_T$ and invariant-mass distributions
of the $q \bar{q} \to Z Z$ process
for $\Lm_\pi=2$, 3, and 5 TeV.
The SM results for $g g \to Z Z$ are plotted
for comparison. } \label{q}
\end{figure}
\end{center}

In Fig.~\ref{q}, we present the RS effects on the process
$q\bar{q} \to Z Z$,
which are determined solely by the $\Lm_\pi$
due to the ignorance of the radion influence.
The KK gravitons can be recognized by
broad peaks.
The RS effects are less important
than those on the $g g \to ZZ$ process:
The effects appear beyond 300 GeV of $p_T$ and
700 GeV of $M_{ZZ}$, generating at most $10^{-3}$ pb/GeV
of the differential cross section with respect to
$p_T$ or $M_{ZZ}$;
the effects on the gluon fusion
shall be shown to appear in the low $p_T$ and $M_{ZZ}$
region where the cross sections are sizable.

\begin{center}
\begin{figure}[htb]
\hbox
to\textwidth{\hss\epsfig{file=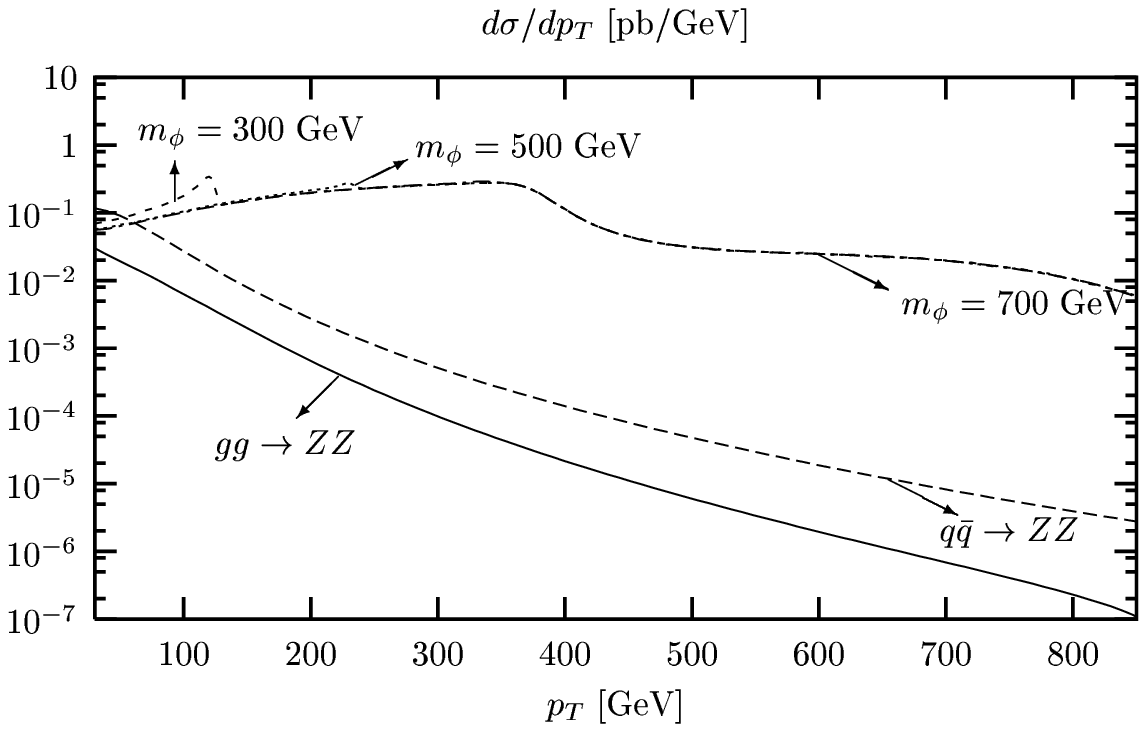,scale=1.1}\hss}
\bigskip
\hbox
to\textwidth{\hss\epsfig{file=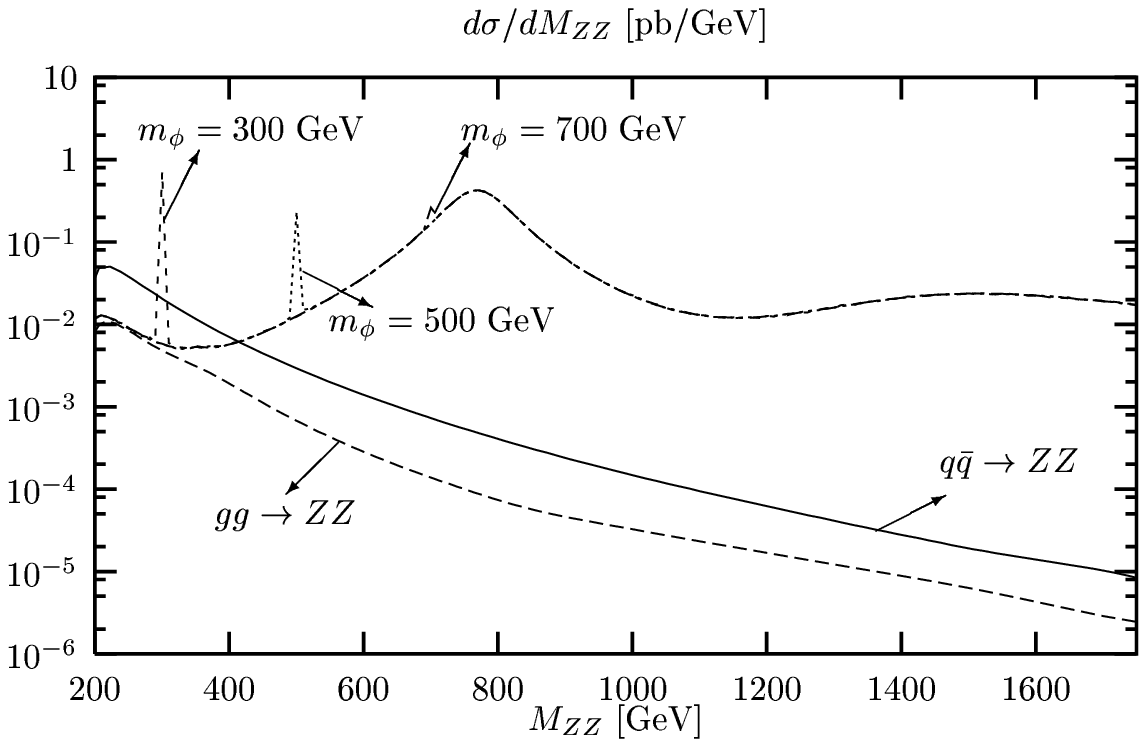,scale=1.1}\hss}
\smallskip
\smallskip
\smallskip
\caption{The $p_T$ and invariant-mass distributions
of the $g g \to Z Z$ process
for $m_\phi=300$, 500, and 700 GeV
with $\Lm_\pi=2$ TeV and $k/M_{\rm Pl}=0.1$.
The SM results for $q\bar{q} \to Z Z$ are plotted
for comparison. } \label{rm}
\end{figure}
\end{center}

\begin{center}
\begin{figure}[htb]
\hbox
to\textwidth{\hss\epsfig{file=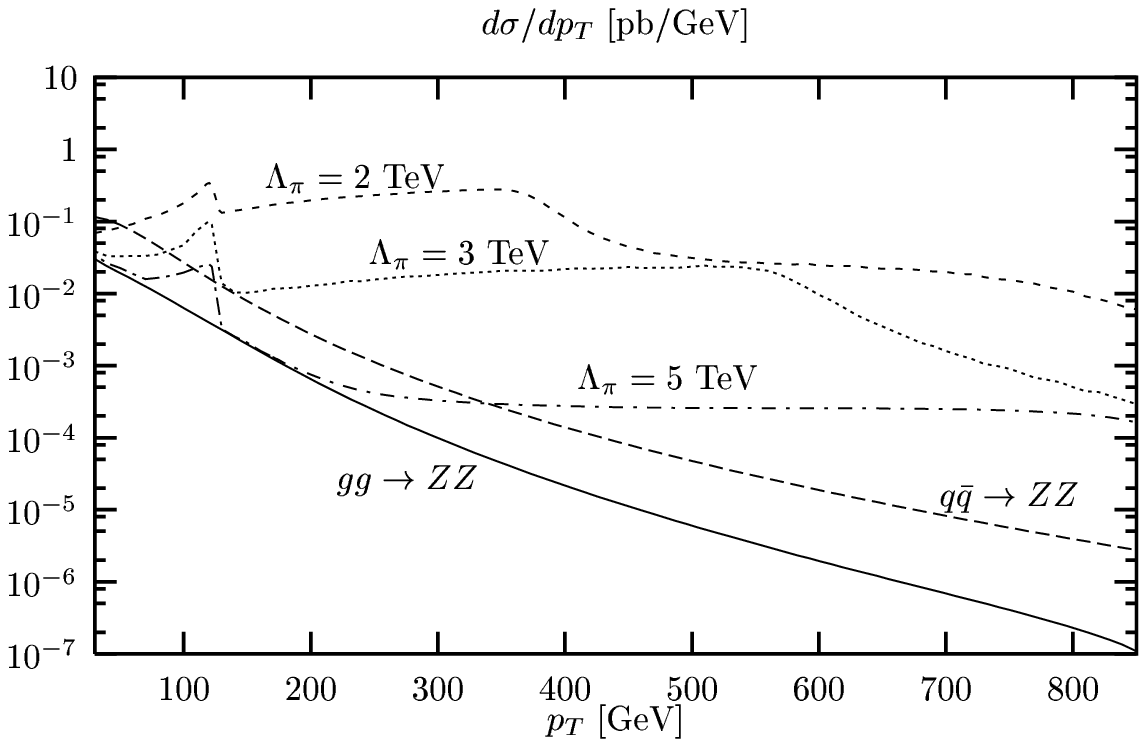,scale=1.1}\hss}
\bigskip
\hbox
to\textwidth{\hss\epsfig{file=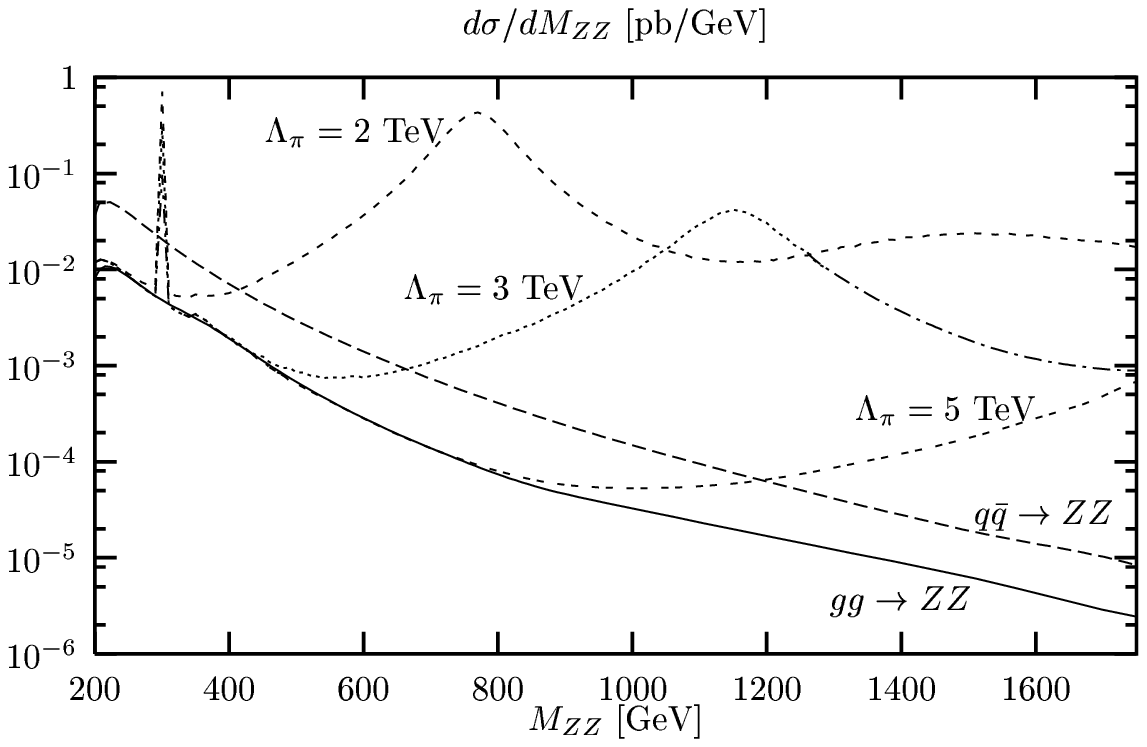,scale=1.1}\hss}
\smallskip
\smallskip
\smallskip
\caption{The $\Lm_\pi$-dependence
on the $p_T$ and invariant-mass distributions
of the $g g\to Z Z$ process
for $\Lm_\pi=2$, 3, and 5 TeV
with $m_\phi=300$ GeV and $k/M_{\rm Pl}=0.1$.} \label{lm}
\end{figure}
\end{center}

\begin{center}
\begin{figure}[htb]
\hbox
to\textwidth{\hss\epsfig{file=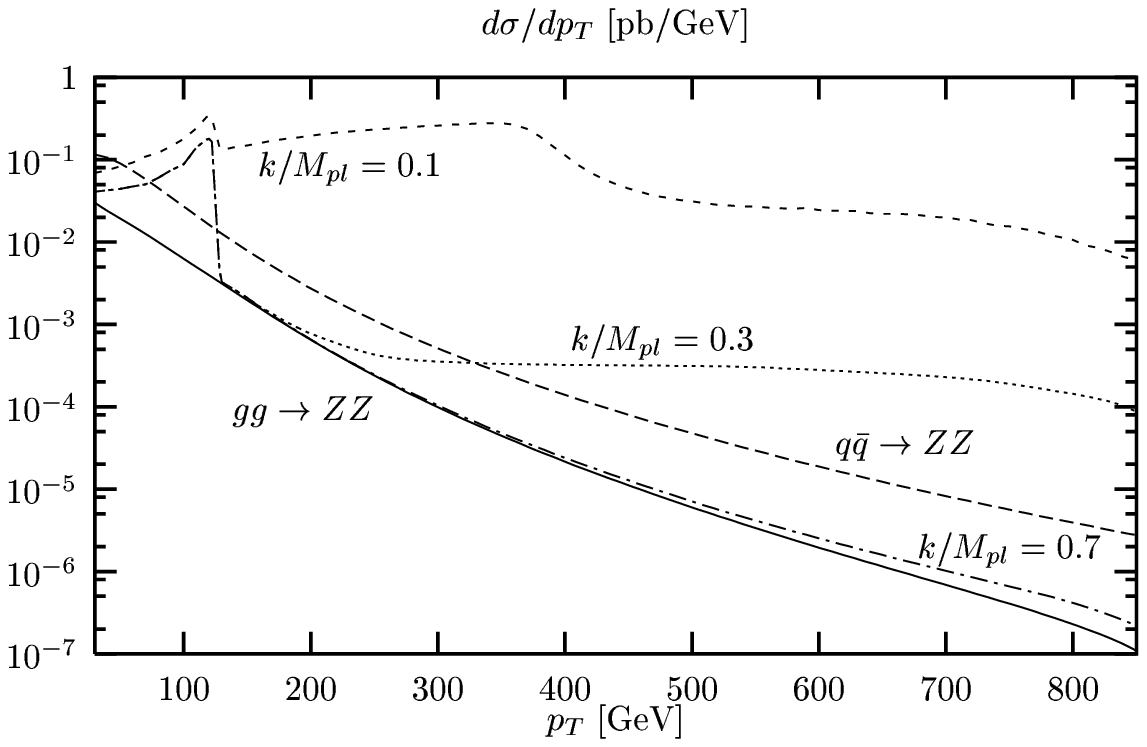,scale=1.1}\hss}
\bigskip
\hbox
to\textwidth{\hss\epsfig{file=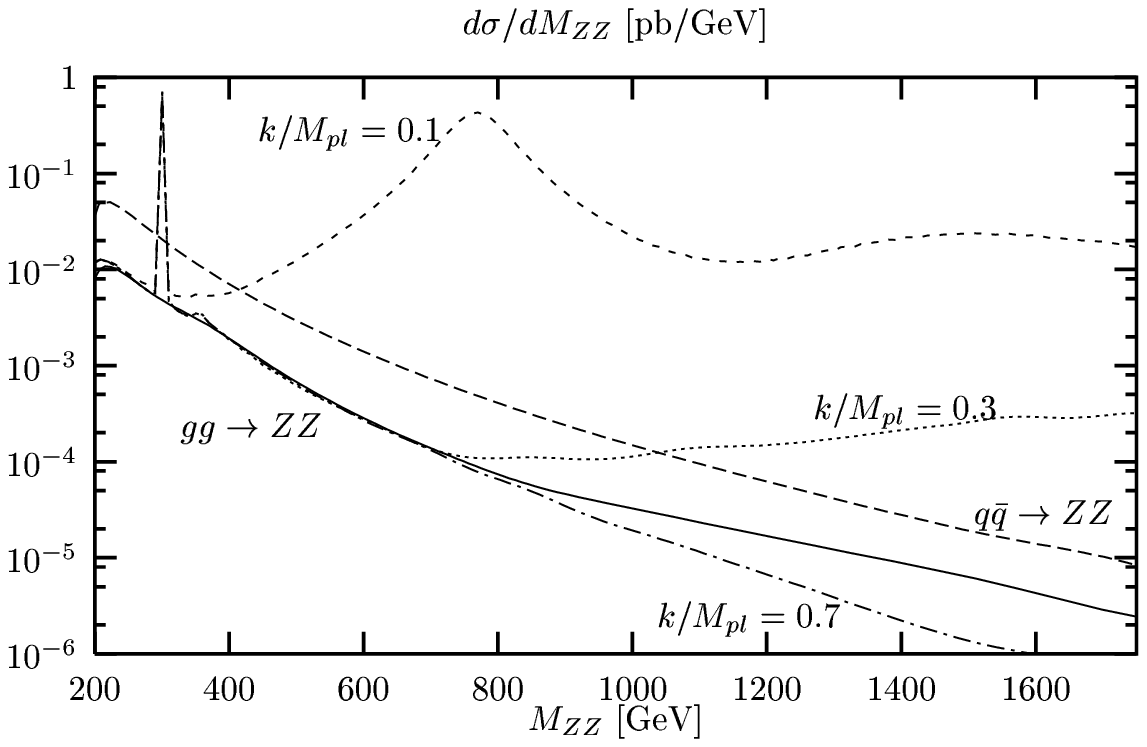,scale=1.1}\hss}
\smallskip
\smallskip
\smallskip
\caption{The $k/M_{\rm Pl}$-dependence
on the $p_T$ and invariant-mass distributions
of the $g g\to Z Z$ process
for $k/M_{\rm Pl}=0.1$, 0.3, and 0.7
with $m_\phi=300$ GeV and $\Lm_\pi=2$ TeV.} \label{k}
\end{figure}
\end{center}

Now we illustrate
each parameter dependence of the RS model to the
gluon fusion process.
In Fig.~\ref{rm},
we plot the distributions
for the $m_\phi=300$, 500, 700 GeV,
with $\Lm_\pi=2$ TeV and $k/M_{\rm Pl}=0.1$.
It can be seen that the resonance peak
of lighter radions is sharper.
If the radion is quite heavy (around 500 GeV
in this parameter space),
the contribution of the KK gravitons to the $p_T$ distributions
overwhelms the radion resonance;
the invariant-mass distribution
is more appropriate to probe.
If the radion is too heavy (more than
700 GeV in this case),
large contributions of the KK gravitons
obscure the radion effects.
Figure \ref{lm} presents the $\Lm_\pi$-dependence
on the $g g\to Z Z$ process
for $\Lm_\pi=2$, 3, and 5 TeV
with $m_\phi=300$ GeV and $k/M_{\rm Pl}=0.1$.
The $M_{ZZ}$ distributions show a sharp resonance
peak of the radion and
the successive broad peaks of the KK gravitons.
In the $p_T$-distributions,
the KK graviton effects yield a plateau region.
The case of $\Lm_\pi \gsim 5$ TeV
would be difficult to probe.
The $k/M_{\rm Pl}$ dependence is presented in Fig.~\ref{k},
for $k/M_{\rm Pl}=0.1$, 0.3, and 0.7
with $m_\phi=300$ GeV and $\Lm_\pi=2$ TeV.
Since the $k/M_{\rm Pl}$ is proportional
to the masses of the KK gravitons (see Eq.~(\ref{Gmass})),
the KK graviton effects
in the large $k/M_{\rm Pl}$ cases are hardly detected.
Note that in the RS model
the magnitude of the five-dimensional curvature
$(R_5=-20\, k^2)$
is required to be smaller than $M_S^2~(\simeq M_{\rm Pl}^2)$
for the reliability of the classical RS solution
derived from the
leading order term in the curvature.
The value of ${k}/{M_{\rm Pl}}$ less than about 0.1 is
theoretically favored \cite{Davoudiasl:2000tf}.

\begin{center}
\begin{figure}[htb]
\hbox
to\textwidth{\hss\epsfig{file=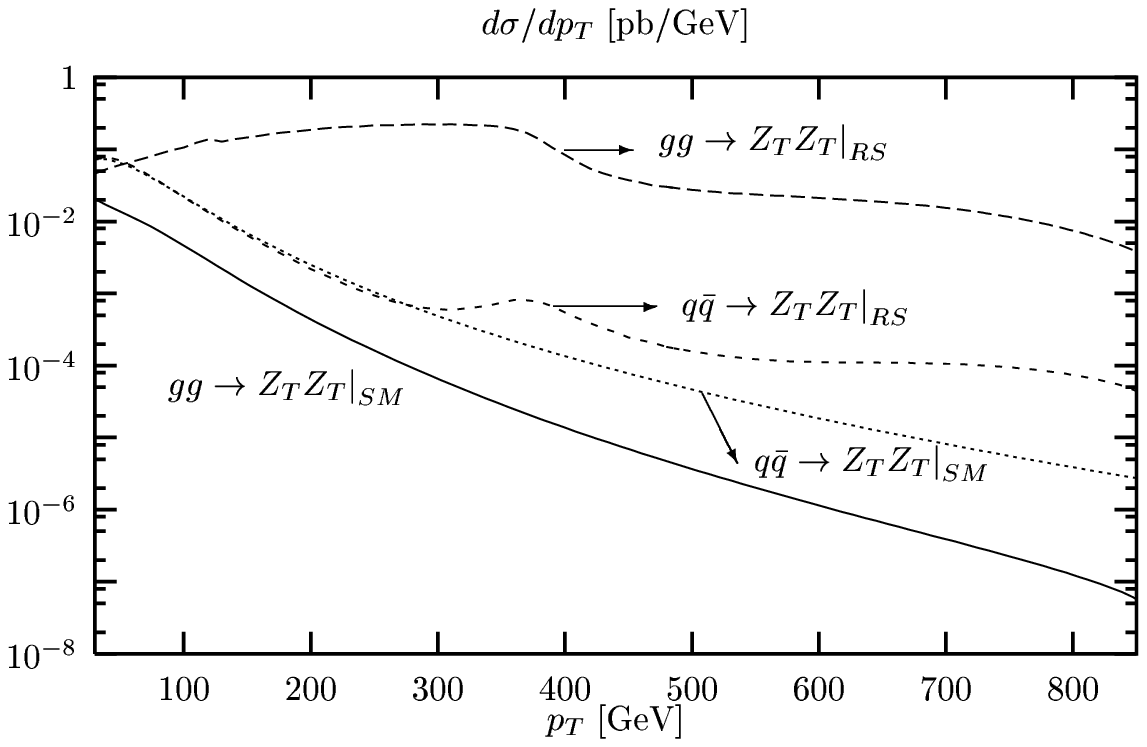,scale=1.1}\hss}
\bigskip
\hbox
to\textwidth{\hss\epsfig{file=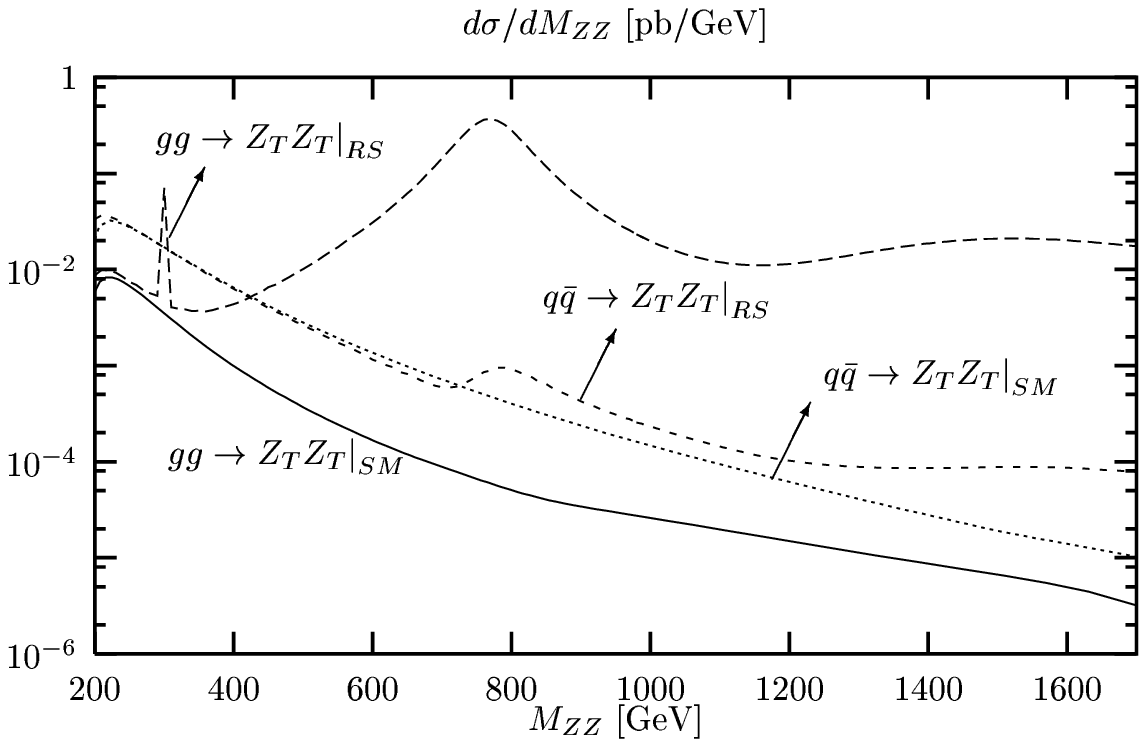,scale=1.1}\hss}
\smallskip
\smallskip
\smallskip
\caption{The $p_T$ and invariant-mass distributions
of the $g g\to Z_T Z_T$ process.
We set $k/M_{\rm Pl}=0.1$, $m_\phi=300$ GeV and
$\Lm_\pi=2$ TeV.} \label{tt}
\end{figure}
\end{center}
\begin{center}
\begin{figure}[htb]
\hbox
to\textwidth{\hss\epsfig{file=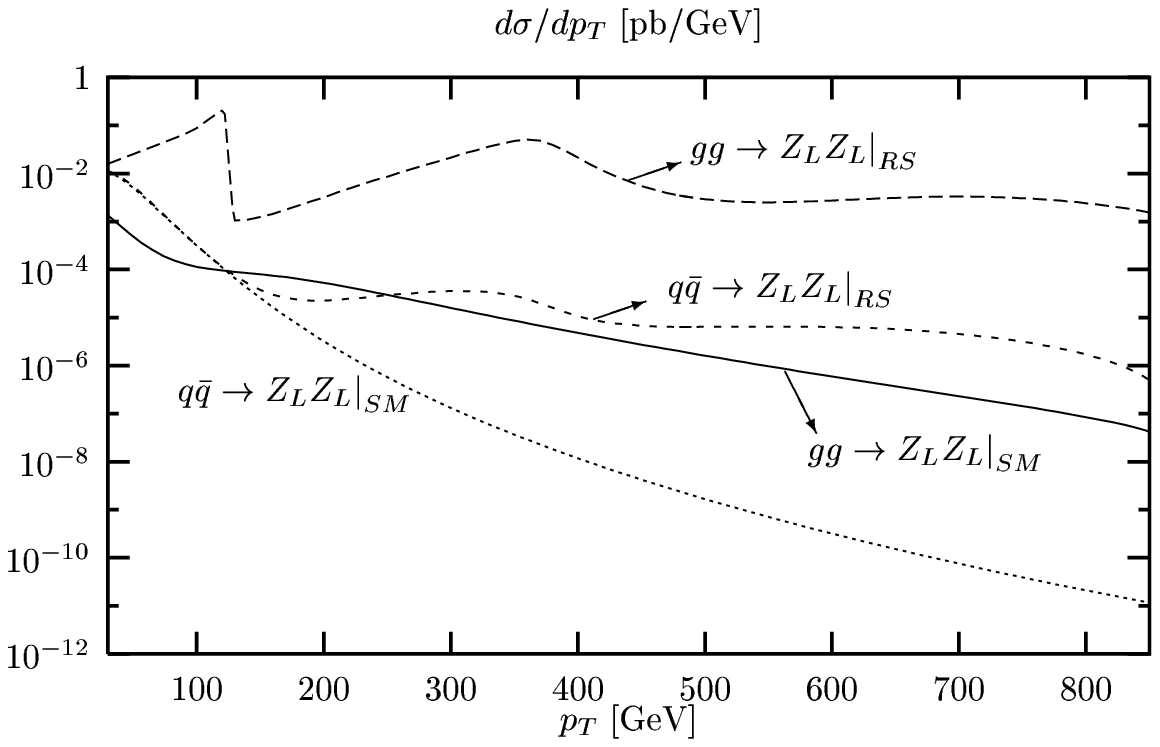,scale=1.1}\hss}
\bigskip
\hbox
to\textwidth{\hss\epsfig{file=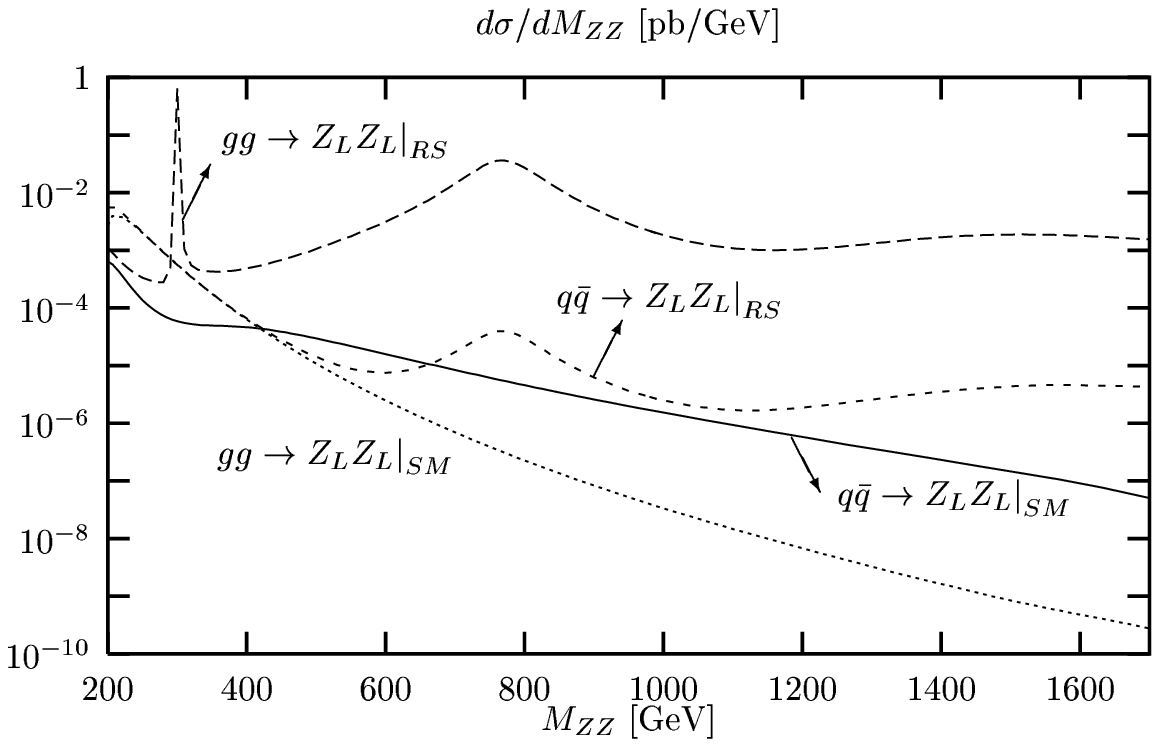,scale=1.1}\hss}
\smallskip
\smallskip
\smallskip
\caption{The $p_T$ and invariant-mass distributions
of the $g g\to Z_L Z_L$ process.
We set $k/M_{\rm Pl}=0.1$, $m_\phi=300$ GeV and
$\Lm_\pi=2$ TeV.} \label{ll}
\end{figure}
\end{center}

We present the influence of the $Z$ polarization
measurement on the distributions.
Figures \ref{tt} and \ref{ll} are
for the polarization states $Z_T Z_T$ and $Z_L Z_L$,
respectively.
We set $\Lambda_\pi =2$ TeV, $k/M_{\rm Pl}=0.1$ and
$m_\phi=300$ GeV for illustration.
As expected from the scalar nature of the radion
and the spin-2 nature of massive KK gravitons,
longitudinally polarized $Z$ bosons
are produced more in the RS
model, while
the SM production of $Z_L Z_L$
through both gluon fusion and $q\bar{q}$ annihilation
is suppressed.
Thus the measurement of the longitudinally polarized
$Z$ bosons would provide
one of the most robust methods
to single out, in particular,
the radion effects of the RS model.

Unfortunately,
an event cannot determine
the polarization of the $Z$ boson.
The angular distributions
for the $Z$ boson decay, $Z \to f\bar{f}$,
provide some information on the $Z$ polarizations.
To leading order,
the RS effects can be ignored in the $Z$ decay.
Neglecting the mass of the final state fermions,
the angular distributions of the $Z$ decay rate
are given by,
in the rest frame of the decaying $Z$,\cite{Mahlon:1998jd}
\be
\frac{1}{\Gm_f}
\frac{d \Gm^\pm}{d \cos \chi }
=
\frac{3}{8}
\left[
\alpha_f ( 1 \mp \cos\chi)^2+(1-\alpha_f)(1\pm\cos\chi)^2
\right]
\ee
for the transversely polarized $Z$ bosons, and
\be
\frac{1}{\Gm_f}
\frac{d \Gm^0}{d \cos \chi }
=
\frac{3}{4}
\sin^2\chi
\ee
for the longitudinally polarized $Z$ bosons.
For the charged leptonic decay,
we have $\alpha_f=(1-2 \,x_W)^2/(1-4x_W
+8x_W^2)$ with $x_W \equiv \sin^2\theta_W$.
The partial width $\Gm_f$ is for the normalization, and
the $\chi$ is the angle between the fermion momentum direction
and the spin axis as seen in the $Z$ rest frame
(the $Z$ boost direction in the helicity basis).
Therefore, an appropriate cut on the $\chi$
would select more data
of longitudinally polarized $Z$ bosons.
We define the ratio $\R$ as follows,
which is proportional to the observable RS effects:
\be
\R \equiv
\frac{
\sg (p p \to Z Z \to l^+ l^- l'^+ l'^-)_{SM+RS}
}{
\sg (p p \to Z Z \to l^+ l^- l'^+ l'^-)_{SM}
}
\,,
\ee
where $l,l'=e,\mu,\tau$.
In order to focus on the detection of the radion effects,
we employ the kinematic cut, e.g., $p_T< 125$~GeV only
for the estimates of the $\R$.
Numerical analysis shows that
with the kinematic cut of $ | \cos\chi |< 0.3$
the $\R$ increases by 49\%
with respect to the $\R$ without any cut on $\chi$.
Thus appropriate kinematic cuts
on the transverse momentum of $Z$ boson
and the angle $\chi$ can enhance
the signal for the radion effects
by, e.g., about 50\%.

\begin{center}
\begin{figure}[htb]
\hbox
to\textwidth{\hss\epsfig{file=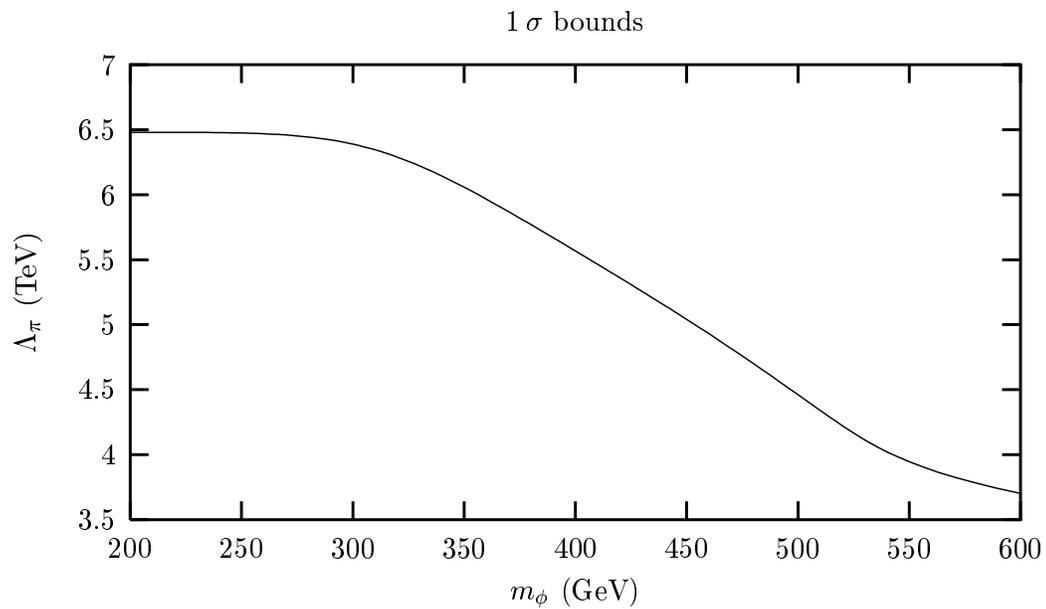,scale=1.1}\hss}
\smallskip
\smallskip
\caption{The $1 \sigma$ sensitivity bounds on
$(\Lm_\pi, m_\phi)$ with $k/M_{\rm Pl}=0.1$.} \label{bound}
\end{figure}
\end{center}

Finally, we estimate the $1\,\sg$ search bounds on the
$\Lm_\pi$ and the radion mass,
which is obtained by comparing
the $total$ cross sections with and without the RS effects:
Even if the cross section of longitudinally polarized $Z$ bosons ($\sg_{LL}$)
and/or the radion resonance peak are powerful methods to probe the RS effects,
higher sensitivity bounds, which is usually relevant in the case of no signal,
are obtained from the observable with $larger$ event number.
As discussed before,
the RS model has unitarity violation
at high energies, which may induce unphysically large
contributions to the total cross section.
One possible way is to exclude data with
high invariant-mass,
i.e.,  to employ an upper cut on the $M_{ZZ}$
such as $\sqrt{\hat{s}} \leq 0.9\; \Lm_\pi$.
Thus we require that
\be
\frac{\sg^<_{_{SM+RS}}-\sg^<_{_{SM}} }{\sqrt{\sg^<_{_{SM}} } }
\sqrt{\mathcal L} \,\epsilon \geq 1,
\ee
where the $\sg^<$ denotes the total
cross section
with an additional kinematic cut of
$
M_{ZZ} < 900\, {\rm GeV}
$.
The $\es$ is the reconstruction efficiency
for the $Z$-boson pair, which is the squared branching ratio
of the $Z$ boson into $e^+ e^-$ or $\mu^+ \mu^-$.
The LHC luminosity ${\mathcal L}$ is 100 pb$^{-1}$.
In Fig.~\ref{bound},
we show the $1\,\sg$ attainable bounds on $\Lm_\pi$ and $m_\phi$
with $k/M_{\rm Pl}=0.1$.
Even with restrictive efficiency,
the $\Lm_\pi$ of about 5 TeV can be experimentally
examined.
Since the radion  has relatively small influence
upon the total cross section,
the $p_T$ or invariant-mass distribution
would be more appropriate to signal the
radion effects.

\section{Summary and Conclusion}

We have studied the $g g \to Z Z$ process
at LHC as a probe of
the Randall-Sundrum scenario
with the Goldberger-Wise stabilization
mechanism.
Even though the process has been regarded
significant in various aspects
(e.g., in examining the Higgs sector),
the main background of the continuum production of
$q\bar{q} \to Z Z$
is known to be dominant
in the SM and MSSM.
It has been shown that
the comprehensive effects of Kaluza-Klein
gravitons and the radion
enhance the chance to probe the model:
The RS effects on
the $q\bar{q} \to Z Z$ process,
which are generated only by KK gravitons,
are much smaller than those on the
$g g \to Z Z$ one;
the KK graviton effects increase
the cross section of $g g \to Z Z$ throughout
the $p_T$ and invariant-mass distributions;
the resonance shape of the radion
is distinguishable from that of the Higgs boson.
Numerical results of the $p_T$ and
invariant-mass distributions
have been obtained
to show the dependence of
the RS model parameters, $(\Lambda_\pi, k/M_{\rm Pl}, m_\phi)$.
The distinction between the Higgs boson and radion
even with the same mass
is feasible since the resonance peak of the radion
is narrower than that of the Higgs boson
in most of the parameter space.
The $p_T$ and invariant-mass distributions
for the polarized $Z$ boson pair have been also
presented.
We have shown that
especially the production of longitudinally polarized
$Z$ bosons,
to which the SM contributions are suppressed,
receives substantial corrections due to KK gravitons
and the radion.
Polarization measurements would provide one of the
most robust methods to signal the RS effects.
The $1\, \sigma$ sensitivity bounds on $(\Lambda_\pi, m_\phi)$
with $k/M_{\rm Pl} =0.1$ have been also obtained.
The $\Lambda_\pi$ of
about $5$ TeV can be experimentally searched even with
restrictive experimental efficiency.
In conclusion,
the channel of $Z$-boson pair production at LHC
with the measurement of the $Z$-polarizations
and the kinematic distributions
can provide
an efficient method to probe the effects of the
RS scenario with the modules fields being stabilized
by the Goldberger-Wise mechanism.

\acknowledgments

\noindent
We would like to thank K.Y. Lee and W.Y. Song
for useful discussions on this work.
The work was supported by the BK21 Program.



\bigskip

\newpage

\newpage

\newpage

\newpage

\newpage

\newpage

\newpage

\newpage

\newpage

\end{document}